\documentstyle[aps,prb,epsf,epsfig]{revtex}

\begin{document}

\twocolumn[\hsize\textwidth\columnwidth\hsize\csname@twocolumnfalse\endcsname

\title{Expansion in the ``distance'' from $H_{c2}(T)$ line for the mixed state of
BCS superconductors}
\author{Anton Knigavko and Frank Marsiglio}
\address{Department of Physics, University of Alberta, Edmonton, Canada T6G 2J1}
\date{\today}
\maketitle

\begin{abstract}
We develop a description of the mixed state of type II superconductivity
valid within a wide range of temperatures and external magnetic fields. 
It is based on the quasiclassical version of microscopic BCS theory and 
employs an expansion in the ``distance'' from the $H_{c2}(T)$ line.
We prove that in a clean metal with spherical Fermi surface and isotropic 
pairing interaction the superconducting condensation always produces 
a hexagonal vortex lattice. 
\end{abstract}
\bigskip

]


The theoretical description of the mixed state is a dif\-ficult problem: the
equations are nonlinear, the solution for the ground state is inhomogeneous
and selfconsistency is required. Analytical progress is facilitated if a
small parameter is present in the theory. In particular, a perturbative
treatment partially removes the need for selfconsistency. Near the second
order phase transition the order parameter is small. The Ginzburg--Landau
(GL) approach is a tool to describe such situations, and it was proven
successful on numerous occasions, including the very discovery of the upper
critical field $H_{c2}$ and the vortex nature of the mixed state\cite
{Abrikosov1}. The $H_{c2}(T)$ line can be calculated from the microscopic
BCS theory of superconductivity: in some symmetric cases exactly \cite
{Helfand1,Scharnberg1} and more often approximately\cite{Scharnberg2,Maki1}.
The framework here is the Gorkov formalism 
or, even more conveniently, the quasiclassical Eilenberger equations. 

The aim of this paper is to develop a rigorous perturbative method to
describe the mixed state. An analytical description would allow to follow
general trends as one moves on $T-H$ plane. To find the structure of a
vortex solid at an arbitrary $(T,H)$ point directly is only possible by
numerical means and constitute demanding work\cite{Klein1,Machida1}.
Therefore, one often starts from the $H_{c2}(T)$ line and exploit in some
way the smallness of the pairing amplitude in its vicinity (see for example 
\cite{Ovchinnikov1}). It is worth emphasizing, however, that the order
parameter is a {\it function} of spatial coordinates while to perform an
expansion unambiguously a small {\it number} is required. Below we argue
that the ``distance'' to the $H_{c2}(T)$ line can be used as such a small
number and we demonstrate how to develop the expansion. We find that this
parameter remains small on a very large portion of the superconducting
region of $T-H$ \ plane, indicating at possible large range of validity of
the expansion. The idea to use the ``distance'' to the $H_{c2}(T)$ line was
recently introduced by B. Rosenstein \cite{Rosenstein1} for the analysis of
fluctuations in the framework of GL theory. For the sake of simplicity we
consider in this paper the case of a clean metal with spherical Fermi
surface (FS) and isotropic pairing interaction. As an immediate result we
are able to prove that superconducting condensation produces the hexagonal
vortex lattice at any temperature. A further development of this formalism
to more complicated cases 
is possible. It could provide, in particular, valuable information on vortex
lattices in unconventional superconductors \cite{VL-sructure-unconv} which
are under scrutiny at present, but have been analyzed mostly in the GL
framework so far.


The Eilenberger equations \ relate the quasiclassical Green functions, both
normal $g(\omega _{n};{\bf v,x})$ and anomalous $f(\omega _{n};{\bf v,x}),$ $%
\bar{f}(\omega _{n};{\bf v,x}),$ and the superconducting order parameter $%
\Delta ({\bf x)}$\cite{Eilenberger1,Larkin1} (see Ref. \cite{Rainer1} for
review). Using the notations ${\bf v}$ for the unit vector in the direction
of the Fermi velocity, $\omega _{n}\equiv 2\pi t(n+1/2)$ for Matsubara
frequencies and$\ D_{i}=\partial _{i}-{\rm i\,}e^{\ast }A_{i}$ for covariant
derivatives, they can be written as 
\begin{eqnarray}
\left( \omega _{n}+{\bf v\,D}/2\right) f &=&\Delta \,g,\;  \label{Eilen-eq-f}
\\
\left( \omega _{n}-{\bf v\,D}^{\ast }/2\right) \bar{f} &=&\Delta ^{\ast }g,
\label{Eilen-eq-fbar} \\
f\;\bar{f}+g^{2} &=&1.  \label{constraint}
\end{eqnarray}
The last relation is the constraint on the quasiclassical Green functions
which can be used instead of the differential equation for $g$. Below we use
the following units: $T_{c}$ for temperature ($t=T/T_{c}$ is reduced
temperature) and the gap function, $R\equiv \hbar v_{F}/T_{c}$ for distances
and $H_{R}\equiv $ $\hbar v_{F}/e^{\ast }R^{2}$ for magnetic field ($%
b=B/H_{R}$ is reduced magnetic induction). The free energy density is given
in units of $N_{0}T_{c}^{2}$ with $N_{0}$ being the density of states of a
normal metal.

To complete the description the selfconsistency equations are required. The
gap equation is used it in the form \cite{Eilenberger1} 
\begin{equation}
\Delta ({\bf x)}\ln \frac{1}{t}=2\pi t\sum_{n=0}^{\infty }\left[ \frac{%
\Delta ({\bf x)}}{\omega _{n}}-\oint_{FS}f(\omega _{n};{\bf v,x})\right] .
\label{gap-eq}
\end{equation}
The strength of the pairing interaction enters Eq. (\ref{gap-eq}) only
implicitly via $T_{c}$. We use spherical angles $(\theta ,\varphi )$ to
specify ${\bf v}.$ Then the FS averaging is simply $\oint_{FS}\equiv \int
\sin \theta {\rm d}\theta {\rm d}\varphi /4\pi $. The other selfconsistency
equation is for supercurrents. It can be disregarded if we concentrate on
strongly type II superconductors with the GL parameter $\kappa \gg 1.$ In
this case the internal magnetic field of a superconductor is constant with
high accuracy, of the order $1/\kappa ^{2}.$ Removal of this approximation
does not invalidate the ideas we discuss below, but adds some complexity to
the formalism and will be considered elsewhere.

We consider the free energy of a superconductor in the form introduced by
Eilenberger\cite{Eilenberger1}. It is a functional over the order parameter $%
\Delta $ as well as the quasiclassical anomalous Green functions $f$ and $%
\bar{f}$ \ (while $g$ is a dependent variable by virtue of Eq. (\ref
{constraint})). The magnetic term $({\bf \nabla \times A})^{2}$ is constant
in the approximation we use and can be omitted. The\ free energy density
reads \cite{Eilenberger1}

\begin{eqnarray}
F &=&\frac{1}{V}\int_{{\bf x}}\left[ |\Delta |^{2}\ln t+2\pi
t\sum_{n=0}^{\infty }\left( \frac{|\Delta |^{2}}{\omega _{n}}-\oint_{FS}%
\left[ \vphantom{\frac{1}{2f}}f\Delta ^{\ast }+\bar{f}\Delta \right. \right.
\right.  \nonumber \\
&&\left. \left. \left. -\left( 1-g\right) \left\{ 2\omega _{n}+\frac{{\bf v}%
}{2f}{\bf D}f-\frac{{\bf v}}{2\bar{f}}{\bf D}^{\ast }\bar{f}\right\} \right]
\right) \right] ,  \label{Eilenberger-energy}
\end{eqnarray}
where the ${\bf x}$ integration is performed over the volume $V$ of the
system. The thermodynamic variables for the free energy is temperature and
magnetic induction, therefore we will discuss $t-b$ plane in what follows.
As usual, $b$ is related to the external magnetic field by thermodynamic\
arguments.

Making use of the smallness of $\Delta $ near $H_{c2}(t)$ line, we proceed
according to the following strategy. First, we solve the Eilenberger
equations for $f$ and $g$ perturbatively in $\Delta $ and obtain the gap
equation solely in terms of $\Delta .$ From that point we deal with this
equation only. The second step is to work out the $H_{c2}$ problem: an
eigenvalue problem given by the linearized gap equation. After $H_{c2}(t)$
line is known the parameter $a_{h}$, which specifies the distance from this
line to an arbitrary $(t,b)$ point, can be defined and the expansion for the
order parameter $\Delta $ itself in terms of $a_{h}$ can be constructed. In
this paper we demonstrate how to obtain the coefficients of the $a_{h}$
expansion in general and calculate them, along with the free energy, to the
lowest order. 
The Green functions are then also known to the same order.


{\it Step 1}. The structure of Eqs. (\ref{Eilen-eq-f})-(\ref{constraint})
suggests the expansions: 
\begin{equation}
g=g_{0}+g_{2}+...,\hspace{0.5cm}f=f_{1}+f_{3}+...,  \label{f-g-expansion}
\end{equation}
and analogously for $\bar{f}.$ Subscripts signify the power of $\Delta $ to
which the term is proportional. The expansions can be worked out easily if
``inversion'' operators $\hat{P}\equiv 1/(\omega _{n}+{\bf v\cdot D}/2)$ and 
$\hat{P}^{\prime }\equiv 1/(\omega _{n}-{\bf v\cdot D}^{\ast }/2)$ are used
to solve\ Eq. (\ref{Eilen-eq-f}) for $f$ and Eq. (\ref{Eilen-eq-fbar}) for $%
\bar{f}$ . Then, starting from the known $g_{0}={\rm sgn}(\omega _{n}),$ we
obtain: 
\begin{eqnarray}
f_{1} &=&g_{0}\hat{P}\Delta ,\hspace{0.5cm}\bar{f}_{1}=g_{0}\hat{P}^{\prime
}\Delta ^{\ast },  \label{E-eq-order-1} \\
g_{2} &=&-\frac{g_{0}}{2}\left( \hat{P}\Delta \right) \left( \hat{P}^{\prime
}\Delta ^{\ast }\right) ,  \label{E-eq-order-2} \\
f_{3} &=&-\frac{g_{0}}{2}\hat{P}\left[ \Delta \left( \hat{P}\Delta \right)
\left( \hat{P}^{\prime }\Delta ^{\ast }\right) \right]  \label{E-eq-order-3}
\end{eqnarray}
and so on. \ The ''inverse'' operators are conveniently defined using an
integral identity: $\hat{P}({\bf v})\equiv \int_{0}^{\infty }{\rm d}\tau
\,\exp \left[ -\tau \left( \left| \omega _{n}\right| +{\bf v\cdot \hat{D}}%
/2\right) \right] $ (see for example discussion in Ref. \cite{inv-operator})
and $\hat{P}^{\prime }({\bf v})=\hat{P}^{\ast }(-{\bf v}).$ Introducing
creation and annihilation operators by $\hat{a}\equiv \Lambda \left( D_{x}+%
{\rm i}D_{y}\right) /\sqrt{2}$, $\hat{a}^{\dagger }\equiv -\Lambda \left(
D_{x}-{\rm i}D_{y}\right) /\sqrt{2}$ where $\Lambda \equiv 1/\sqrt{b}$ is
the magnetic length we arrive at 
\begin{eqnarray}
\hat{P} &=&\int_{0}^{\infty }{\rm d}\tau \,{\rm e}^{-\tau |\text{\thinspace }%
\omega _{n}|-\tau ^{2}\left( \frac{\sin \theta }{4\Lambda }\right)
^{2}}\sum_{m_{1}m_{2}=0}^{\infty }{\rm e}^{{\rm i}(m_{1}-m_{2})\varphi } 
\nonumber \\
&&\times \left( \frac{\tau \sin \theta }{2\sqrt{2}\Lambda }\right)
^{m_{1}+m_{2}}\frac{\left( -1\right) ^{m_{2}}}{m_{1}!m_{2}!}\left( \hat{a}%
^{\dagger }\right) ^{m_{1}}\hat{a}^{m_{2}}  \label{P-operator}
\end{eqnarray}
It was assumed that nothing in our problem depends on $x_{3},$ the distance
along external magnetic field direction (we are dealing with the ground
state).

{\it Step 2.} The linearized gap equation is obtained by substituting $f_{1}$
from Eq. (\ref{E-eq-order-1}) in Eq. (\ref{gap-eq}). The $\varphi $
integration produces $m_{1}=m_{2}$ in the $\hat{P}$ operator 
meaning that only the number operator $\hat{a}^{\dagger }\hat{%
a}$ and its powers are left and, consequently, the eigenfunctions are the
Landau levels. The highest critical temperature at a given external magnetic
field is achieved for zero Landau level: $\Delta ({\bf x)}\sim \psi _{0}(%
{\bf x})$.\cite{Helfand1} Note that in this case only the first term with $%
m_{1}=m_{2}=0$ in Eq. (\ref{P-operator}), which does not depend on
spatial derivatives, contributes to
$H_{c2}$.

{\it Step 3.} We are now in a position to introduce the actual small
parameter $a_{h}$ controlling the problem. For this purpose we consider
again the term with $f_{1}$ in the gap equation Eq. (\ref{gap-eq}) and,
following the idea of Ref. \cite{Rosenstein1}, separate the constant part
with $m=0$ of the $\hat{P}$ operator  from Eq. (\ref{P-operator}). Then the
gap equation can be written as follows: 
\begin{equation}
{\cal H}\Delta =a_{h}\Delta +2\pi t\sum\nolimits_{n=0}^{\infty
}\oint\nolimits_{FS}\left( f_{3}+f_{5}+...\right) ,
\label{gap-eq-restricted}
\end{equation}
where 
\begin{eqnarray}
a_{h} &\equiv &\ln \frac{1}{t}-2\pi t\sum_{n=0}^{\infty }\left[ \frac{1}{%
\omega _{n}}-\int_{0}^{\pi }\sin \theta \frac{{\rm d}\theta }{2}\right. 
\nonumber \\
&&\left. \times 
\displaystyle\int %
_{0}^{\infty }{\rm d}\tau \,e^{-\tau \text{\thinspace }\omega _{n}-\tau
^{2}\left( \frac{\sin \theta }{4\Lambda }\right) ^{2}}\right] ,
\label{a-parameter} \\
{\cal H} &\equiv &-2\pi t\sum_{n=0}^{\infty }\int_{0}^{\pi }\sin \theta 
\frac{{\rm d}\theta }{2}%
\displaystyle\int %
_{0}^{\infty }{\rm d}\tau \,e^{-\tau \text{\thinspace }\omega _{n}-\tau
^{2}\left( \frac{\sin \theta }{4\Lambda }\right) ^{2}}  \nonumber \\
&&\times \sum_{m=1}^{\infty }\left( \tau \frac{\sin \theta }{4\Lambda }%
\right) ^{2m}\frac{\left( -2\right) ^{m}}{\left( m!\right) ^{2}}\left( \hat{a%
}^{\dagger }\right) ^{m}\hat{a}^{m}  \label{H-operator}
\end{eqnarray}
and $f_{3},$ $f_{5},$ $...$ have to be taken from Eqs. (\ref{E-eq-order-1}%
)--(\ref{E-eq-order-3}), etc. As discussed above, the $H_{c2}(t)$ line is
given by $a_{h}(t,b)=0$ which is identical to the result of Helfand and
Werthamer \cite{Helfand1}. For an arbitrary $(t,b)$ point the parameter $%
a_{h}$ specifies its ``closeness'' to the $H_{c2}(t)$ line. It is clear that 
$a_{h}$ can be used to develop an expansion for the function $\Delta ({\bf x}%
)$ in the vicinity of the $H_{c2}(t)$ line where $a_{h}\ll 1$ by
construction. Eq. (\ref{gap-eq-restricted}) dictates: 
\begin{equation}
\Delta =\sqrt{a_{h}}\left( \Delta _{0}+a_{h}\Delta _{1}+a_{h}^{2}\Delta
_{2}+...\right)  \label{Exp-for Delta}
\end{equation}
with the structure identical to that found by the $a_{h}$ expansion of GL
theory \cite{Rosenstein1} and constitute, most probably, a consequence of
the fact that the effective electron--electron interaction is treated in the
BCS theory of superconductivity in the saddle point approximation. The
expansion of Eq. (\ref{Exp-for Delta}) could work well quite far away from $%
H_{c2}(t).$ To test this assumption we computed the function $a_{h}(t,b)$.
As seen in the insert in Fig. 1 the necessary condition for convergence, $%
a_{h}<1$, is fulfilled in a very large portion of the $t-b$ plane. 

By construction the operator ${\cal H}$ defined by Eq. (\ref{H-operator})
does not contain a constant term. This facilitates all manipulations with
Eq. (\ref{gap-eq-restricted}). In particular, it is easily checked that $%
\Delta _{0}$ contains the zero Landau level only since ${\cal H}\Delta
_{0}=0.$ The equations in the next two orders read: 
\begin{eqnarray}
{\cal H}\Delta _{1} &=&\Delta _{0}-2\pi t\sum_{n=0}^{\infty
}\oint\nolimits_{FS}\frac{1}{2}\hat{P}\left[ \Delta _{0}(\hat{P}\Delta _{0})(%
\hat{P}^{\prime }\Delta _{0}^{\ast })\right] ,  \label{order-1-eq} \\
{\cal H}\Delta _{2} &=&\Delta _{1}+2\pi t\sum_{n=0}^{\infty
}\oint\nolimits_{FS}\left[ f_{3}\left\{ \Delta _{0}^{2}\Delta _{1}\right\}
+f_{5}\left\{ \Delta _{0}^{5}\right\} \right] .  \label{order-2-eq}
\end{eqnarray}
The form of the quantities appearing in the integrand of Eq. (\ref
{order-2-eq}) is self-explanatory. For example, $f_{3}\left\{ \Delta
_{0}^{2}\Delta _{1}\right\} $ is a sum of the expressions given by Eq. (\ref
{E-eq-order-3}) with $\Delta $ substituted for by $\Delta _{0}$ twice and by 
$\Delta _{1}$ once in all possible combination. We do not write down the
lengthy exact expressions since we do not use them below. We observe that
the expansion in powers of $a_{h}$ leads to the mixing of orders of the
preliminary expansion of Green function in terms of $\Delta $ (see Eq. (\ref
{f-g-expansion})). Also, the higher order is a term in the $a_{h}$ expansion
for $\Delta ,$ Eq. (\ref{Exp-for Delta}), the larger is the maximal power of
nonlinear contributions to it.

{\it Step 4. }We look for solutions for every $\Delta _{i}$ in the form of
expansions in the Landau levels' basis $\left\{ \psi _{m}\right\} $: 
\begin{equation}
\Delta _{0}=c\,\psi _{0},\hspace{0.5cm}\Delta
_{1}=\sum\nolimits_{m=0}^{\infty }c_{1m}\psi _{m},\hspace{0.5cm}...
\label{Form-of-Delta}
\end{equation}
In the complete basis of Landau levels there exists another parameter,
besides $m$, for labeling each function in the set. It is analogous to wave
vector $k$. Functions with $k\neq 0$ do not contribute to $\Delta $ for
equilibrium solutions studied in this paper, but of course would be relevant
for fluctuations\cite{Rosenstein1}. The structure of the $a_{h}$ expansion
is such that to find the coefficient of $\psi _{0}$ in each order the next
order equation is needed. In particular, the coefficients $c$ and $c_{1m}$
with $m\neq 0$ are found from Eq. (\ref{order-1-eq}) while the coefficients $%
c_{10}$ and $c_{2m}$ with $m\neq 0$ are found from Eq. (\ref{order-2-eq}),
and so on.


{\it \ } Having completed the construction of the $a_{h}$ expansion we
proceed to work it out to the lowest order: to calculate the coefficient $c$
(see Eq. (\ref{Form-of-Delta})). We multiply Eq. (\ref{order-1-eq}) by $\psi
_{0}^{\ast }$ on the left, integrate over ${\bf x}$ and use the
orthonormality of Landau level basis. Then, the term with ${\cal H}$
vanishes while the first term on the left hand side reduces to a constant
and we obtain $c^{-2}=\pi t\sum_{n=0}^{\infty }\oint_{FS}\int_{{\bf x}}\psi
_{0}^{\ast }\hat{P}[\psi _{0}(\hat{P}\psi _{0})(\hat{P}^{\prime }\psi
_{0}^{\ast })]$. At this stage the ideas developed for GL theory in Ref. 
\cite{Saint-James} turn out to be very effective. In order to lower the free
energy$\ $the coordinate dependence of the order parameter$\ \Delta ({\bf x})
$ should be that of a periodic lattice. This can be parametrized by just a
few numbers. They are used to minimize the free energy calculated first with
a lattice of an arbitrary shape. Such a minimization procedure is a
necessary step for the $a_{h}$ expansion too, but we emphasize that this
nonperturbative step is required only once, in the lowest order. 
We consider Eq. (\ref{Eilenberger-energy}), and constraining $\Delta $ 
and $f$ to satisfy the Euler equation of this functional, 
Eqs. (\ref{Eilen-eq-f})--(\ref{gap-eq}), we obtain the equilibrium free 
energy density: 
\begin{equation}
F_{equil}=-\frac{1}{V}\int\limits_{{\bf x}}2\pi t\sum_{n=0}^{\infty
}\oint\limits_{FS}\frac{1-g}{1+g}\frac{\Delta ^{\ast }({\bf x)}f+\Delta (%
{\bf x)}\bar{f}}{2}.  \label{Energy-transformed}
\end{equation}
We check that, when $\Delta $ is small, $F_{equil}$ starts from the order $%
\Delta ^{4}$ (see Eqs. (\ref{E-eq-order-1})--(\ref{E-eq-order-2}) and note
that $g_{0}=1$ for $n>0$). The order $\Delta ^{2}$ has vanished, as it
should be. Next we insert the $a_{h}$ expansion for the order parameter $%
\Delta ({\bf x})$ and obtain $F_{equil}=-c^{4}a_{h}^{2}\,\frac{\pi t}{4}%
\sum_{n}\oint_{FS}\int_{{\bf x}}[\psi _{0}^{\ast }(\hat{P}^{\prime }\psi
_{0}^{\ast })(\hat{P}\psi _{0})^{2}+cc]$ in the lowest order, which is
needed for the minimization. The set $\left\{ \psi _{m}\right\} $ of lattice
shaped Landau levels we used is described in detail in Ref. \cite
{Zhitomirsky1}. For parameters specifying the geometry of the lattice we
made a conventional choice of $\rho =(a_{2}/a_{1})\cos \alpha $ and $\sigma
=(a_{2}/a_{1})\sin \alpha $ where $a_{1}$, $a_{2}$ are edges of the unit
cell parallelogram and $\alpha $ is the angle in between. In the isotropic
case two parameters suffice because the global orientation of the lattice is
irrelevant while the area of the unit cell is fixed by the flux quantization.
The further calculations for both the free energy $F_{equil}$ and the
coefficient $c$ contain similar steps. After the explicit form of $\hat{P}$
operator is inserted into Eqs. (\ref{order-1-eq}) and (\ref
{Energy-transformed}) the $\varphi $ integration and the $\omega _{n}$
summation can be performed. Handling numerous $m$ summations and $\tau $
integrals allows to present the free energy density, in units of $%
N_{0}T_{c}^{2},$ as 
\begin{equation}
F_{equil}=-\frac{1}{2}\frac{a_{h}^{2}(t,b)}{\beta _{E}(\rho ,\sigma ;t,b)},
\label{Free-energy-main}
\end{equation}
where the notation $\beta _{E}\equiv c^{-2}$ was introduced. Subscript ``E''
stands for Eilenberger. Eq. (\ref{Free-energy-main}) can be viewed as a
generalization of the corresponding result of GL theory, which has the same
form but with Abrikosov's energy parameter $\beta _{A}=\int_{{\bf x}}\left|
\psi _{0}\right| ^{4}$ instead of $\beta _{E}$. Eilenberger's energy
parameter is given by: 
\begin{equation}
\beta _{E}=\frac{1}{2\sqrt{2}\,t}\sum_{m_{1}m_{2}=0}^{\infty
}C_{m_{1}m_{2}}(d)\,\beta _{m_{1}m_{2}}(\rho ,\sigma ).  \label{beta-Eilen}
\end{equation}
Temperature and magnetic induction enter the sums in Eq. (\ref{beta-Eilen})
via the combination $d\equiv t/\sqrt{b}$ only. The coefficients $%
C_{m_{1}m_{2}}$ read: 
\begin{eqnarray*}
&&C_{m_{1}m_{2}}=\int\limits_{-1}^{1}\frac{{\rm d}s{\rm d}y{\rm d}z}{8}%
\int\limits_{0}^{\infty }{\rm d}u\frac{u\,{\rm e}^{-\,Ku^{2}}}{\sinh u}\,\,%
\frac{\left( 1+z\right) }{\sqrt{1+z^{2}}}\left( 1-s\right) ^{m_{1}} \\
&&\times \left( 1+s\right) ^{m_{2}}\left[ \frac{\left( 1-y^{2}\right) \left(
1-z^{2}\right) }{2}\left( \frac{u}{4d}\right) ^{2}\right] ^{m_{1}+m_{2}}.
\end{eqnarray*}
The infinite $u$ integral is cut off either by the Gaussian or by $u/\sinh u$%
. The range of the former depends on $d$ because 
\[
K=\frac{1-y^{2}}{\left( 4d\right) ^{2}}\left[ (1-z)^{2}+\frac{1+s^{2}}{2}%
(1+z)^{2}\right] .
\]
As a result, two different limiting regimes exist for the energy parameter: $%
\beta _{E}(t,b)\sim 1/\sqrt{b}$ if $d\ll 1$ and $\beta _{E}(t,b)\sim 1/t$ if 
$d\gg 1.$ All geometrical information about the lattice is contained in
matrices $\beta _{m_{1}m_{2}}=\int_{{\bf x}}\psi _{0}\psi _{m_{1}+m_{2}}\psi
_{m_{1}}^{\ast }\psi _{m_{2}}^{\ast }\ $which can be cast in the form: 
\begin{eqnarray*}
&&\beta _{m_{1}m_{2}}=\frac{\sqrt{\sigma }}{\left( -2\right) ^{m_{1}+m_{2}}}%
\sum_{l_{1}l_{2}=-\infty }^{+\infty }\cos \left[ 2\pi \rho l_{1}l_{2}\right]
\,{\rm e}^{-\pi \sigma \left( l_{1}^{2}+l_{2}^{2}\right) } \\
&&\times
\sum_{k_{1}=0}^{m_{1}}\sum_{k_{2}=0}^{m_{2}}\sum_{k_{3}=0}^{k_{1}+k_{2}}2^{%
\frac{k_{1}+k_{2}+k_{3}}{2}}\frac{H_{m_{1}-k_{1}}\left[ \sqrt{\pi \sigma }%
\left( l_{2}-l_{1}\right) \right] }{\left( \frac{k_{3}-k_{1}+k_{2}}{2}%
\right) !\left( m_{1}-k_{1}\right) !} \\
&&\times \frac{H_{m_{2}-k_{2}}\left[ \sqrt{\pi \sigma }\left(
l_{1}-l_{2}\right) \right] \,H_{m_{1}+m_{2}-k_{3}}\left[ \sqrt{\pi \sigma }%
\left( l_{1}+l_{2}\right) \right] }{\left( \frac{k_{3}+k_{1}-k_{2}}{2}%
\right) !\left( m_{2}-k_{2}\right) !\left( \frac{k_{1}+k_{2}-k_{3}}{2}%
\right) !\left( m_{1}+m_{2}-k_{3}\right) !},
\end{eqnarray*}
where $H_{m}[...]$ is Hermite polynomial.

To find the global minimum of $F_{equil}$ from Eq. (\ref{Free-energy-main})
we evaluated $\beta _{E}(\rho ,\sigma )$ along the boundary of the
fundamental domain for these geometrical parameters\cite{Zhitomirsky1}.\
This boundary corresponds to all possible two fold symmetric lattices and
this is the location of extrema. We considered $(t,b)$ values belonging the $%
H_{c2}(t)$ line and found that the minimum of $\beta _{E}$ is always at $%
\rho =1/2$, $\sigma =\sqrt{3}/2$, that is the hexagonal lattice. 
Along the $H_{c2}(t)$ line $\beta_{E}(1/2,\sqrt{3}/2)$ changes only slightly:
from 0.90 at t=1 to 1.23 at t=0 (see Fig.1). 
Note that it is finite at $t=0$ even though the $\hat{P}$ operator,
Eq. (\ref{P-operator}), is poorly defined at this temperature. The
calculation of $\beta _{E}$ along the $H_{c2}(t)$ line produces also the
values of $\beta _{E}$ for all $(t,b)$ points below this line because $\beta
_{E}$ is a function of $d=t/\sqrt{b}$ only, disregarding the  $1/t$ factor
common to all terms in Eq. (\ref{beta-Eilen}). 

The GL results appear from microscopic expressions in the limit 
$d\rightarrow \infty$ (or $t\rightarrow 1,b\rightarrow 0$). 
This limit produces only an asymptotic series, not a divergent 
one\cite{Helfand1}. However, the linear $H_{c2}(t)$
curve, standard in GL theory and obtainable just with the first two 
terms of the $1/d$ expansion in the equation $a_{h}=0$ (see Eq. (\ref
{a-parameter})), is in an excellent agreement with the exact $H_{c2}(t)$ line
down to as low as $t=0.5$. Similar situation turns out to occur for the free
energy. Our analysis has revealed that even at low temperatures  the first 
term with $m_{1}=m_{2}=0$ in Eq. (\ref{beta-Eilen}) exceeds the sum over 
all Landau levels by about 15\% only (see Fig.1). 
This interesting fact could be one of the reasons why GL theory enjoyed 
so many successes while used very far beyond the domain of its strict 
validity.


In conclusion, we developed a microscopic description of type II
superconductivity based on the expansion in ``distance'' from the $H_{c2}(T)$
line, the $a_{h}$ expansion. The isotropic case was treated in detail as an
example. Anisotropic FS and/or pairing interaction would produce a different 
$H_{c2}(T)$ line. Still, as long as it can be found exactly the $a_{h}$
expansion is possible to construct in the same manner in the entire $T-H$
plane. If an approximate expression for the $H_{c2}(T)$ line is used as a
starting point, the resulting $a_{h}$ expansion would have a smaller range
of applicability. The presented method seems to be practical because the
actual shape of FS can be conveniently modelled in the framework of the
quasiclassical approach.

This research was supported by NSERC and CIAR.

\begin{figure}[bth]
\epsfig{figure=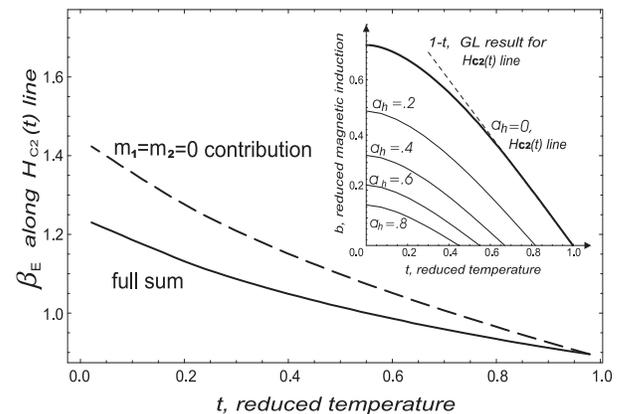,height=5.5cm,width=8.0cm}
\caption{Energy parameter $\protect\beta_E$ for the hexagonal lattice 
along $H_{c2}(t)$ line presented as a function of temperature: 
dashed line --- the contribution of the first term in 
Eq. (\ref{beta-Eilen}) with $m_1=m_2=0$, 
solid line --- the full sum.   
Insert: Contour plot of the parameter $a_h$ as a function of $t$ 
and $b$. Magnetic induction is normalized to produce 
${\rm d}b/{\rm d}t=-1$ at $t=1$.}
\label{fig1}
\end{figure}

\end{document}